\documentclass{ws-ijmpd}
\usepackage[super,compress]{cite}
\begin{document}

\markboth{Authors' Names}
{Instructions for Typing Manuscripts (Paper's Title)}

%
\catchline{}{}{}{}{}
%

\title{UPDATE ON ELECTROMAGNETICALLY ACCELERATING UNIVERSE}

\author{PAUL H FRAMPTON}

\address{Dipartimento di Matematica e Fisica "Ennio de Giorgio",\\
Universita del Salento and INFN-Lecce,\\
Via Arnesano,\\ 73100 Lecce,
Italy.\\
paul.h.frampton@gmail.com}

\author{Talk at the 17th Marcel Grossmann Meeting, Pescara, Italy. \\
July 9th 2024, to appear in the Proceedings.} 

\maketitle

\begin{history}
\received{Day Month Year}
\revised{Day Month Year}
\end{history}

\begin{abstract}
We discuss a cosmological model first proposed in 2022
in which the accelerated rate of expansion is attributed to
repulsive electromagnetic forces between like-sign charged
PEMNSs (= Primordial Extremely Massive Naked Singularities).
\end{abstract}


\ccode{PACS numbers:}

\newpage

\noindent
\section{Introduction to the EAU Model} 

\bigskip

\noindent
Theoretical cosmology is at an exciting stage because about 95\% of the energy in
the Visible Universe remains incompletely understood. The 25\% which
is Dark Matter has constituents whose mass is unknown by over one hundred
orders of magnitude. The 70\% which is Dark Energy is, if anything, more mysterious.
Although it can be parametrised by a Cosmological Constant with equation of state
$\omega = -1$ which provides an excellent phenomenological description,
that is only a parametrisation and not a complete understanding.

\bigskip

\noindent
In this talk, we address the issues of Dark Matter and Dark Energy
using a novel approach\cite{1,2,3}. We use only the classical theories of
electrodynamics and general relativity.
We shall not employ any knowledge of quantum mechanics or of
theories describing the short range strong and weak interactions.

\bigskip

\noindent
This talk may be regarded as a follow up to a 2018 paper
entitled {\it On the Origin and Nature of Dark
Matter} and could have simply added {\it and Energy}
to that title. We have, however, chosen {\it Update on Electromagnetically
Accelerating Universe} because it more accurately characterises
our present emphasis on the EAU model whose main idea is that electromagnetism
dominates over gravitation in the explanation of the accelerating
cosmological expansion. This idea takes us beyond the first paper (Einstein)
which applied general relativity to theoretical cosmology.
This is not surprising, since in 1917 that author was obviously unaware
of the fact discovered only in 1998 that the rate of cosmological expansion
is accelerating.

\bigskip

\noindent
The make up of this talk is that Primordial Black Holes
are discussed  in Section 2, then Primordial Naked
Singularities in Section 3. Finally in 
Section 4 there is a Discussion.

\bigskip

\noindent
\section{ Primordial Black Holes (PBHs)}

\bigskip

\noindent
Black holes may be classified into those which arise from the
gravitational collapse of stars and others which do not. By primordial, we shall
refer to all of the others. In general, PBHs with masses up to $10^5 M_{\odot}$
are expected to be formed during the first second after the Big Bang and
arise from inhomogeneities and fluctuations of spacetime. The existence
of PBHs was first proposed by Novikov and Zeldovich
and independently seven years later in the West by Carr and Hawking
The idea that the dark matter
constituents are PBHs was first suggested by Chapline.

\newpage

\noindent
Shortly after the original presentation of general relativity
a metric describing a static black hole of mass M with zero charge
and zero spin was discovered by Schwarzschild
in the form

\begin{equation}
ds^2 = - \left( 1 - \frac{r_S}{r} \right) dt^2 + \left( 1 - \frac{r_S}{r} \right)^{-1} dr^2 + r^2 d\Omega^2
\label{Schwarzschild}
\end{equation}

\noindent
Shortly thereafter, the Reissner-Nordstrom metric
for a static Black Hole with electric charge was found. It then took 
a surprising forty-five years
until Kerr cleverly found a solution of general relativity
corresponding to a such a solution with spin. We shall not
discuss the case of non-zero spin in the present paper because,
although we expect that all the objects we discuss do spin in
Nature, according to the calculations which
use Kerr's generalisation, spin is an inessential 
complication in all of our subsequent considerations.

\bigskip

\subsection{Primordial Intermediate Mass Black\\
Holes as Galactic Dark Matter}

\bigskip

\noindent
Global fits to cosmological parameters have led to a consensus that about one quarter
of the energy of the universe is in the form of 
electrically-neutral dark matter. It seemed natural to propose
 that the dark matter constituents are primordial black holes with masses many
times the mass of the Sun. In a galaxy like the Milky Way, the proposal was that
residing in the galaxy are between ten million and ten billion black holes with masses between
one hundred and one hundred thousand solar masses.

\bigskip

\noindent
Black holes in this range of masses are naturally known as Intermediate Mass Black Hole (IMBHs)
since they lie intermediate between the masses of stellar-mass black holes
and the masses of the
supermassive black holes at galactic centres.

\bigskip

\noindent
The existence of stellar mass black holes in Nature was established sixty
years ago in 1964 by the discovery in Cygnus X-1 of such a black hole
with mass about $15M_{\odot}$. Such X-Ray binaries  were studied 
in which black holes
appear in the mass range between 
$5M_{\odot}$ and  $100M_{\odot}$,

\bigskip

\noindent
The existence of dark matter was first discovered by 
Zwicky in 1933
in the Coma Cluster. Its presence in individual galaxies was demonstrated
convincingly by Rubin in the 1970s 
from measurement of the rotation curves which demanded the
existence of additional matter to what was luminous.  

\bigskip

\noindent
The PBH mass function is all important,
Possible PBH masses extend upwards to many solar
masses and without any obvious upper limit, far beyond
what was was thought possible in the twentieth century when ignorance about
PBHs with many solar masses probably prevented the MACHO
and EROS Collaborations from 
discovering a larger fraction of the dark matter.

\bigskip

\noindent
Black holes formed by gravitational collapse cannot satisfy
$M_{BH} \ll M_{\odot}$ because stars powered by nuclear fusion
cannot be far below $M = M_{\odot}$.
This was contradicted by the studies in primordiality
which suggested that much lighter black holes can be produced 
in the earliest stages of the Big Bang.

\bigskip

\noindent
Such PBHs are of special interest for several reasons. Firstly, they are the only type
of black hole which can be so light, down to $10^{12}kg\sim10^{-18}M_{\odot}$, that Hawking radiation might conceivably be detected. Secondly, PBHs in the intermediate-mass region
$100M_{\odot} \leq M_{IMBH} \leq 10^5M_{\odot}$ can provide the galactic
dark matter. 

\bigskip

\noindent
The mechanism of PBH formation involves large fluctuations or inhomogeneities. 
Carr and Hawking
argued that we know there are fluctuations in the universe in order to seed
structure formation and there must similarly be fluctuations in the early universe. Provided the
radiation is compressed to a high density, meaning to a radius as small
as its Schwarzschild radius, a PBH will form. Because the density in the early
universe is extremely high, it is very likely that PBHs will be created. The two
necessities are high density which is guaranteed and large inhomogeneities
which are possible.

\bigskip

\noindent
During radiation domination
\begin{equation}
a(t) \propto t^{1/2}
\label{raddom}
\end{equation}

\noindent
and 

\begin{equation}
\rho_{\gamma} \propto a(t)^{-4} \propto t^{-2}
\label{rhogamma}
\end{equation}

\bigskip

\noindent
Ignoring factors $O(1)$ and bearing in
mind that the radius of a black hole is 

\begin{equation}
r_{BH} \sim \left( \frac{M_{BH}}{M_{Planck}^2} \right)
\label{BHradius}
\end{equation}

\noindent
with

\begin{equation}
M_{Planck} \sim 10^{-8} kg \sim 10^{-38}M_{\odot}.
\label{MPlanck}
\end{equation} 

\noindent
Let us define a Planck density $\rho_{Planck}$  by

\begin{equation}
\rho_{Planck} \sim (10^{-5}g)(10^{-33}cm)^{-3} = 10^{94} \rho_{H_2 O}.
\label{rhoPlanck} 
\end{equation}

\bigskip

\noindent
The density of a black hole $\rho_{BH}(M_{BH})$ is

\begin{eqnarray}
\rho_{BH}(M_{BH}) &\sim& \left( \frac{M_{BH}}{r_{BH}^3} \right) \nonumber\\
&= &\rho_{Planck} \left(\frac{M_{Planck}}{M_{BH}} \right)^2 \nonumber \\
&\sim& 10^{94} \rho_{H_2 O} \left( \frac{10^{-38} M_{\odot}}{M_{BH}} \right)^2 \nonumber \\
\label{rhoBH}
\end{eqnarray}

\noindent
which means that for a solar-mass black hole

\begin{equation}
\rho_{BH}(M_{\odot}) \sim 10^{18} \rho_{H_2 O}
\label{rhoBHSun}
\end{equation}

\noindent
while for a billion solar mass black hole

\begin{equation}
\rho_{BH}(10^9M_{\odot}) \sim \rho_{H_2 O}.
\label{rhoBHSM}
\end{equation}

\noindent
and above this mass the density falls as $M_{BH}^{-2}$.

\bigskip

\noindent
The mass of the PBH is derived by combining
Eqs. (\ref{rhogamma}) and (\ref{rhoBH}). We see from these 
two equations that $M_{PBH}$ grows linearly with time and using  
Solar Mass units we find 

\begin{equation}
M_{PBH} \sim \left( \frac{t}{1 sec} \right) 10^5 M_{\odot}
\label{PBHmass}
\end{equation}

\noindent
which implies, if we insist on PBH formation
before the electroweak phase transition, $t < 10^{-12}s$, that

\begin{equation}
M_{PBH} < 10^{-7} M_{\odot}
\label{upperbound}
\end{equation}

\bigskip

\noindent
Such an upper bound as Eq.(\ref{upperbound}) explains why the MACHO
searches at the turn of the twenty-first century, 
inspired by the clever  suggestion of Paczynski, lacked motivation
to pursue searching above $100M_{\odot}$ because it was thought 
incorrectly at that
time that PBHs were far too light. It was known correctly
that the results of gravitational collapse of
normal stars, or even large early stars, were below $100M_{\odot}$.
Supermassive black holes with $M > 10^6M_{\odot}$
such as $Sag A^*$ in the Milky Way were beginning to be discovered in galactic centers
but their origin was unclear and will be
discussed further in Section 2.2. 

\bigskip

\noindent
Using the mechanism for Hawking radiation provides the lifetime for a 
black hole evaporating 
{\it in vacuo} given by

\begin{equation}
\tau_{BH} \sim \left( \frac{M_{BH}}{M_{\odot}} \right)^3 \times 10^{64} years
\label{BHlifetime}
\end{equation}

\noindent
so that to survive for the age $10^{10}$ years of the universe, there is a
lower bound on $M_{PBH}$ to augment the upper bound in Eq.(\ref{upperbound}),
giving as the full range of Carr-Hawking PBHs:

\begin{equation}
10^{-18}M_{\odot} < M_{PBH} < 10^{-7} M_{\odot}
\label{CarrHawking}
\end{equation}

\noindent
The lowest mass possible for s surviving PBH in Eq.(\ref{CarrHawking}) has the density
$\rho \sim 10^{58} \rho_{H_2 O}$. It is an object which has the physical size of a proton and
the mass of Mount Everest !!

\bigskip

\noindent
The Hawking temperature $T_H(M_{BH})$
of a black hole is given by

\begin{equation}
T_H(M_{BH}) = 6 \times 10^{-8}K \left( \frac{M_{\odot}}{M_{BH}} \right)
\label{Hawking}
\end{equation}

\noindent
which would be above the CMB temperature, and hence there would be outgoing radiation for all of the cases with $M_{BH} < 2 \times 10^{-8}M_{\odot}$.
Hypothetically, if the dark matter halo were made entirely of the brightest possible (in terms of Hawking radiation) $10^{-18}M_{\odot}$
PBHs, the expected distance to the nearest PBH would be about $10^7$ km.
Although the PBH temperature, according to Eq. (\ref{Hawking}) is $\sim 6\times 10^{10} K$,
the inverse square law 
renders the intensity of Hawking radiation too small, by many orders of magnitude,
to allow detection by any foreseeable terrestrial apparatus.

\bigskip

\noindent
The originally suggested mechanism produces PBHs with masses in the range
up to $10^{-7} M_{\odot}$. We shall now discuss formation of far more
massive PBHs
by a quite different mechanism.
As already discussed, PBH formation requires very large inhomogeneities.
Here we shall illustrate how mathematically to produce inhomogeneities
which are exponentially large.

\bigskip

\noindent
In the simplest single-stage inflation, no exceptionally large density 
perturbation is expected. Therefore it is necessary to consider
at least a two-stage hybrid inflation with respective fields called
inflaton and waterfall. The idea then involves parametric
resonance in that, after the first of the two stages of inflation, mutual couplings of the
inflaton and waterfall fields cause both to oscillate arbitraily wildly
and produce perturbations which can grow exponentially.

\bigskip

\noindent
A second (waterfall) inflation then stretches further the inhomogeneities,
thus enabling production of PBHs with arbitrarily high mass.
This specific model may not describe Nature
but provides an existence theorem to confirm that
arbitrarily large mass PBHs can be produced mathematically. 
The resulting mass function
is spiked, but it is possible that other PBH production mechanisms can
produce a smoother mass function.

\bigskip

\noindent
Full details of the model are presented in 2010 (F.K.T.Y) where the inflaton
and waterfall fields are denoted by $\sigma$ and $\psi$ respectively.
Between the two stages of inflation, the $\sigma$ and $\psi$ fields oscillate, decaying into their quanta via their self and mutual couplings. Specific modes of $\sigma$ and $\psi$ are amplified by
parametric resonance. The resulting coupled equations for the 
two fields are of Mathieu type 
with the exponentially-growing solutions. 

\bigskip

\noindent
Numerical
solution shows that the peak wave number $k_{peak}$ is approximately linear in
$m_{\sigma}$. The resultant PBH mass, the horizon 
mass when the fluctuations re-enter the horizon, is approximately

\begin{equation}
M_{PBH} \sim 1.4 \times 10^{13}M_{\odot} \left( \frac{k_{peak}}{Mpc^{-1}} \right)^{-2}
\label{PBHmass}
\end{equation}

\bigskip

\noindent
Explicit plots were exhibited in  FKTY 2010 
for the cases $M_{PBH} = 10^{-8}M_{\odot},
 10^{-7}M_{\odot}$ and $10^5M_{\odot}$.
At that time (2010) , although not included in the paper, it was confirmed that parameters 
can always be chosen such that arbitrarily high mass 
PBHs, at or even beyond the mass of the universe, may
be produced. This is an important result to be borne in mind.

\bigskip

\noindent
In the PBH production mechanism based on hybrid inflation with parametric 
resonance, the mass function is generally sharply spiked at a specific mass region.
Such a peculiar mass function is not expected to be a general feature of PBH formation, 
only a property of this specific mechanism.
But this specific mechanism readily demonstrates the possibility of primordial formation of black holes
with many solar masses. For completeness, it should be pointed out that PBHs with
masses up to $10^{-15}M_{\odot}$ were discussed already in the 1970s,
for example by Carr and by Novikov, Polnarev, Starobinskii
and Zeldovich.

\bigskip

\noindent
For dark matter in galaxies, PIMBHs are important,
 where the upper end must be truncated at 
$10^{5}M_{\odot}$ to stay well away 
from galactic disk instability first discussed by Ostriker {\it et al}.
They showed  convincingly that an object with mass one million solar
masses out in the spiral arms of the Milky Way destabilizes the
galactic disk to such an extent that the entire galaxy collapses.

\bigskip

\noindent
Observations of rotation curves reveal that the dark matter in galaxies
including the Milky Way fills out an approximately spherical halo
somewhat larger in radius than the Disk occupied by the luminous stars.
Numerical simulations of structure formation suggest a profile of the
dark matter of the NFW type.

\bigskip

\noindent
Note that he NFW profile is 
independent of the mass of the dark matter constituent. and the
numerical calculations are restricted by the available computer size, 
for a system as large as a typical galaxy, to constituents which
have many solar masses.

\bigskip

\noindent
In our discussion of 2015, we focused on 
galaxies like the Milky Way
and restricted the mass range for the dark matter constituents 
to lie within the three orders of magnitude

\begin{equation}
10^2 M_{\odot} < M < 10^5M_{\odot}
\label{Frampton}
\end{equation}

\noindent
We shall not repeat lengthy entropy arguments here,
 just to say that the constituents were proposed to be Primordial Intermediate 
 Mass Black Holes, PIMBHs. 
 
 \bigskip
 
 \noindent
Assuming a
total dark halo mass of $10^{12}M_{\odot}$, Eq.(\ref{Frampton}) implies
that the number $N$ of PIMBHs
is between ten million ($10^7$) and ten billion ($10^{10}$) 
Assuming further that the dark halo has a radius $R$ of a 
hundred thousand ($10^5$)
light years
the mean separation $\bar{L}$ of PIMBHs can then estimated by

\begin{equation}
\bar{L} \sim \left( \frac{R}{N^{1/3} }\right)
\label{meanL}
\end{equation}

\noindent
which translates approximately to
\begin{equation}
100 ly < \bar{L} < 1000 ly
\label{meanL2}
\end{equation}

\noindent
which provides also a reasonable estimate of the distance to the nearest PIMBH
from the Earth which is very far outside the Solar System where the orbital radius
of the outermost planet Neptune is $\sim 0.001$ ly.

\bigskip

\noindent
To an outsider, It may be surprising that millions of intermediate-mass black holes
in the Milky Way can have remained undetected. Ironically, hey could have been
detected more than two decades ago had the MACHO Collaboration persisted in its microlensing experiment at Mount Stromlo
Observatory in Australia.

\bigskip

\noindent
Dark matter was first discovered almost a century ago
by Zwicky in the Coma cluster, a large cluster at 99 Mpc
containing over a thousand galaxies and with
total mass estimated at $6 \times 10^{14}M_{\odot}$.
A convincing proof of the existence 
of cluster dark matter was provided by the Bullet cluster
collision where the distinct behaviours of the X-ray emitting gas which
collides, and the dark matter which does not, was observable.

\bigskip

\noindent
Since there is not the same Disk stability limit as for galaxies,
the constituents of the cluster dark matter can involve also PSMBHs 
up to much higher masses than possible for the PIMBHs within
galaxies

\bigskip

\noindent
The possible solution of the galactic dark matter problem cries out for experimental
verification. Three methods have been discussed: wide binaries,
distortion of the CMB, and microlensing. Of these, microlensing seems the most
direct and the promising.
Microlensing experiments were carried out by the MACHO
and EROS Collaborations decades ago. At that time, it was
believed that PBH masses were below $10^{-7}M_{\odot}$ by virtue of the
Carr-Hawking mechanism. Heavier black holes could, it was then believed,
arise only from gravitational collapse of normal stars, or 
heavier early stars, and would have mass below $100M_{\odot}$. 

\bigskip

\noindent
For this reason, there was no motivation to suspect that there might be
MACHOs which led to higher duration microlensing events.
The longevity, $\hat{t}$, of an event is

\begin{equation}
\hat{t} = 0.2 yrs \left( \frac{M_{PBH}}{M_{\odot}} \right)^{\frac{1}{2}}
\label{longevity}
\end{equation}

\noindent
which assumes a transit velocity $200 km/s$.
Subsituting our extended PBH masses, one finds approximately
$\hat{t} \sim 6, 20, 60$ years for $M_{PBH} \sim 10^3, 10^4, 10^5 M_{\odot}$
respectively.
It is to be hoped that MACHO searches will soon resume 
at the Vera Rubin Observatory and
focus on highest longevity microlensing events.
Is it possible that convincing observations showing only a fraction of a light curve
could suffice? If so, only a fraction of the {\it e.g.} six years, corresponding to PIMBHs with
one thousand solar masses, could be enough to confirm the theory.

\bigskip

\noindent
\subsection{Primordial Supermassive Black Holes \\
(PSMBHs) at Galactic Centers}

\bigskip

\noindent
Evidence for supermassive black holes at galactic centres arises from the
observations of fast-moving stars around them and such stars being swallowed or 
torn apart by the strong gravitational field. The first discovered SMBH was 
the one, Sag $A^*$, at the core of the Milky Way which was discovered
in 1974 and has mass $M_{SagA*} \sim 4.1 \times 10^6M_{\odot}$.
The SMBH at the core of 
the nearby Andromeda galaxy ($M31$)
has mass $M=2\times 10^8M_{\odot}$, fifty times $M_{SagA*}$.
The most massive core SMBH so far observed is for NGC4889 with 
$M \sim 2.1\times 10^9M_{\odot}$. Some galaxies contain two SMBHs in a binary,
expec5ed to be the result of a galaxy merger. 
Quasars contain black holes with even higher masses up to at least
$4\times 10^{10}M_{\odot}$. .

\bigskip

\noindent
A black hole with the mass of $Sag A^*$ would disrupt the disk
dynamics were it out in the spiral arms but at, or
near to, the
center of mass of the Milky Way it is more stable. $Sag A^*$ is far too massive to
have been the result of a gravitational collapse, and if we take
the view that all black holes either are 
the result of gravitational collapse or are primordial then the
galaxies' core SMBHs must be primordial.
Nevertheless, it is probable that the PSMBHs are
built up by merging and accretion from less
massive PIMBH seeds.

\bigskip

\noindent
\section{Primordial Naked Singularities (PNSs)}

\bigskip

\noindent
Just as neutral black holes can be formed as PBHs in the early
universe, it is natrual to assume that objects can be formed based on the
Reissner-Nordstrom metric 

\begin{equation}
ds^2 = f(r) dt^2 - f(r)^{-1}dr^2 - r^2 d\theta^2 -r^2 \sin^2 \theta d\phi^2
\label{RNmetric}
\end{equation}

\noindent
where

\begin{equation}
f(r) \equiv \left( 1 - \frac{r_S}{r} + \frac{r_Q^2}{r^2} \right).
\label{f(r)}
\end{equation}

\noindent
with

\begin{equation}
r_S =   2 G M ~~~~~~ r_Q= Q^2 G
\label{rSrQ}
\end{equation}

\bigskip

\noindent
The horizon(s) of the RN metric occur when

\begin{equation}
f(r) =0
\end{equation}

\noindent
which gives

\begin{equation}
r_{\pm} = \frac{1}{2} \left( r_S \pm \sqrt{r_S^2 - 4 r_Q^2} \right)
\label{rpm}
\end{equation}

\bigskip

\noindent
It follows that for $2r_Q < r_S$, $Q^2 < M$, there are two horizons. 
On the other hand, when $2r_Q = r_S$,
$Q^2=M$ 
the RN black hole is named extremal and there is only one horizon.
If $2r_Q > r_S$,  $Q^2 > M$, the  RN metric may be called super-extremal.
in this case there is no horizon
at all  and the $r=0$ singularity becomes observable to a distant observer. This is called
 a naked singularity. With this last inequality, it is no longer a black hole which, by definition, requires an horizon.

\bigskip

\noindent
Consider two identical objects with mass M and charge Q. Then
the electromagnetic repulsive force $F_{em} \propto k_eQ^2$
and the gravitational attraction $F_{grav} \propto GM^2$.
Thus, for the electromagnetic repulsion to exceed the gravitational
attraction we need $Q^2 > GM^2/k_e$ and hence
perhaps super-extremal Reissner-Nordstrom or Naked
Singularities(NSs)\footnote{To anticipate NSs we shall replace
BH by NS for charged dark matter. If charges satisfy $Q^2<M$
this replacement is unnecessary.} We cannot claim to understand the formation
of PNSs. One idea hinted at in Araya et al.(2022) is that extremely massive
ones, charged PEMNSs might begin life as electrically neutral PBHs
then, during the dark ages, selectively accrete electrons over protons. However this
formation process evolves, it must be completed
before the onset of accelerated expansion some 4 billlion
years ago at cosmic time $t \sim 9.8$ Gy.

\bigskip

\noindent
\subsection{Primordial Extremely Massive\\
Naked Singularities --  the EAU Model}

\bigskip

\noindent
A novel EAU model has been suggested in where
 dark energy is replaced by charged dark matter in the form of PEMNSs or charged
Primordial Extremely Massive Naked Singularities\footnote{In 
the original paper the PEMNSs were called PEMBHs}. That discussion 
involved the new idea that, at the very largest cosmological distances, the dominant force
is electromagnetism rather than gravitation. This differs from the assumption tacitly
made in the first application of general relativity to cosmology by Einstein.

\bigskip

\noindent
The production mechanism for PBHs in general is not well understood, and for the
PEMNSs we shall make the assumption that they are formed
before the accelerated expansion begins at $t=t_{DE}\sim 9.8$ Gy, For the expansion before $t_{DE}$ we shall assume
that the $\Lambda CDM$ model is approximately accurate.

\bigskip

\noindent
The subsequent expansion in the charged dark matter model will in the future
depart markedly from the $\Lambda CDM$ case. We can regard this as advantageous
because the future fate of the universe in the conventional picture does have certain
unaesthetic features in terms of the extremely large size of the asymptotic
extroverse.

\bigskip

\noindent
In the $\Lambda CDM$  model the introverse, or what is also called the visible 
universe, coincides with the extroverse at $t=t_{DE} \sim 9.8$ Gy with the common
radius

\begin{equation}
R_{EV}(t_{DE}) = R_{IV}(t_{DE}) =  39 Gly.
\label{tDE}
\end{equation}

\bigskip

\noindent
The introverse expansion is limited by the speed of light and its radius increases
from Eq. (\ref{tDE}) to 44 Gly at the present time $t=t_0$ but asymptotes only to

\begin{equation}
R_{IV} (t \rightarrow \infty) \rightarrow 58 Gly
\label{RIVasymp}
\end{equation}

\bigskip

\noindent
The extroverse expansion is, by contrast, exponential and superluminal. Its radius increases
from its value 39 Gly in Eq. (\ref{tDE}) to 52 Gly at the present time $t=t_0$ and grows without limit. After only a trillion years it attains the extremely large value

\begin{equation}
R_{EV} (t  = 1 Ty) = 9.7 \times 10^{32} Gly.
\label{REVtrillion}
\end{equation}

\bigskip

\noindent
This future for the $\Lambda CDM$ scenario seems distasteful because the
introverse becomes of ever decreasing, and eventually vanishing, significance,
relative to the extroverse.

\bigskip

\noindent
One attempt at a possible formation mechanism of PEMNSs was provided
by Chileans, Araja {\it et al} where their common sign of electric charge, negative,
arises from preferential accretion of electrons relative to protons. 

\bigskip

\noindent
This
formation mechanism is not well understood
\footnote{Electrically neutral PEMBHS were first considered, with a different acronym SLABs,  by Carr {\it et al}.}
so to create a cosmological model we shall for simplicity assume that
the PEMNSs are all formed before $t=t_{DE} \sim 9.8$ Gy and
thereafter the
Friedmann equation ignoring radiation, is

\begin{equation}
\left( \frac{\dot{a}}{a} \right)^2= \frac{\Lambda(t)}{3} + \frac {8\pi G}{3} \rho_{matter}
\label{Friedmann}
\end{equation}

\noindent
where $\Lambda(t)$ is the cosmological "constant" generated by Coulomb
repulsion between the PEMNSs.
From Eq.(\ref{Friedmann}), in the $\Lambda CDM$ model with $a(t_0) = 1$ and constant $\Lambda(t)\equiv \Lambda_0$, we
would predict that, in the distant future

\begin{equation}
a(t \rightarrow \infty) \sim exp \left( \sqrt{ \frac{\Lambda_0}{3}} (t-t_0) \right)
\label{exponential}
\end{equation}

\bigskip

\noindent
In the case of charged dark matter, with no dark energy, we must
re-write Eq.({\ref{Friedmann}) as

\begin{equation}
\left( \frac{\dot{a}}{a} \right)^2= \frac {8\pi G}{3} \rho_{cPEMNSs}+ \frac {8\pi G}{3} \rho_{matter}
\label{FriedmannPrime}
\end{equation}

\noindent
in which

\begin{equation}
\rho_{matter} (t)  = \frac{\rho_{matter} (t_0)}{a(t)^3}
\label{rhomatter}
\end{equation}

\noindent
where matter includes normal matter and uncharged dark matter.

\bigskip

\noindent
Of special interest to the present discussion is the expected future behaviour
of the charged dark matter

\begin{equation}
\rho_{PEMNSs} (t) = \frac{\rho_{PEMNSs} (t_0)}{a(t)^3}
\label{rhocPEMBHs}
\end{equation}

\noindent
so that comparison of Eq.(\ref{Friedmann}) and Eq.(\ref{FriedmannPrime}) suggests
that the cosmological constant is predicted to decrease from its present value.

\bigskip

\noindent
More specifically, we find
that asymptotically  the scale factor will behave as if matter-dominated
and the cosmological constant will decrease at large future times as a power

\begin{equation}
a(t\rightarrow \infty) \sim t^{\frac{2}{3}} ~~~~~~ \Lambda(t \rightarrow \infty) \sim t^{-2}.
\label{scale}
\end{equation} 

\bigskip

\noindent
so that a trillion years in the future $\Lambda(t)$ will have decreased
by some four orders of magnitude relative to $\Lambda(t_0)$. See
Table 1 {\it ut supra}.

\begin{table}[t]
\caption{COSMOLOGICAL ``CONSTANT".}
\begin{center}
\begin{tabular}{||c|c||}
\hline
\hline
time & $\Lambda(t)$ \\
\hline
\hline
$t_0$ & $(2.0meV)^4$ \\
\hline
\hline
$t_0+10Gy$ & $(1.0 meV)^4$ \\
\hline
\hline
$t_0+100Gy$ & $(700 \mu eV)^4$ \\
\hline
\hline
$t_0+1Ty$ & $(230 \mu eV)^4$ \\
\hline
\hline
$t_0+1Py$ & $(7.4\mu eV)^4$ \\
\hline
\hline
\end{tabular}
\end{center}
\end{table}

\bigskip

\noindent
In both the $\Lambda CDM$ model and the EAU model, the present time
is an unusual time in cosmic history. In the former case, there is the
present similarity between the the densities of dark matter and energy.
In the latter case with charged dark matter, the present accelerated
expansion is maximal and will disappear within a few more billion years.

\bigskip

\noindent
In the EAU model, acceleration 
began about 4 Gy ago at $t_{DE}= 9.8Gy = t_0-4 Gy$.
This behaviour will disappear in a few more billion
years. The value of the cosmological constant is predicted to fall
like $a(t)^{-2}$ so that, when $t \sim \sqrt{2} t_0 \sim 19.5 Gy
\sim t_0 +4.7 Gy$, the value of
$\Lambda(t)$ will be one half of its present value, $\Lambda(t_0)$. 
On the other hand, the equation of state associated 
with $\Lambda$ is predicted to be
accurately $\omega = -1$, so close to that value that measuring the
difference seems forever impracticable.

\bigskip

\noindent
For charged dark matter, we now discuss the future time evolution of the introverse and
extroverse.  For the introverse,
nothing changes from the $\Lambda CDM$, and after a trillion years, the introverse radius
will be at its asymptotic value $R_{IV}=58 Gly$, as stated in Eq.(\ref{RIVasymp}).
By contrast, the future for the extroverse
is very different for charged dark matter than for the conventional $\Lambda CDM$ case.
WIth the growth $a(t) \propto t^{\frac{2}{3}}$ we find
that the radius of the extroverse at $t=1$ Ty is 

\begin{equation}
R_{EV}(t=1Ty) \sim 900 Gly.
\label{REVnew}
\end{equation}

\bigskip
\noindent
This is in stark contrast to the extremely large value
$9.7 \times 10^{32}$ Gly predicted by the $\Lambda CDM$ model, quoted in
Eq.(\ref{REVtrillion}) above. Eq.(\ref{REVnew}) means that if there still exist
scientific observers their view
of the distant universe will be quite similar  to the present one and will
include many billions of galaxies.

\bigskip

\noindent
In the $\Lambda CDM$ case, such a hypothetical observational cosmologist, trillions of years
in the future,
could observe only the Milky Way and objects which are gravitationally bound to it, so that cosmology would become
an extinct science.

\bigskip

\noindent
The principal physics advantage of charged dark matter is that it avoids the 
idea of an unknown repulsive gravity inherent in ''dark energy". Electromagnetism
provides the only known long-range repulsion so it is more attractive
to adopt it as the explanation for the accelerating universe.
The secondary advantage of charged dark matter, that it provides a conducive 
environment for observational cosmology trillions of years into the
future, is not by itself a sufficient reason to select a theory.

\bigskip

\noindent
\section{Discussion}

\bigskip
\bigskip

\noindent
Although this talk is essentially speculative, we are unaware
of any fatal flaw. We have replaced the conventional make up
for the slices of the universe's energy pie (5\% normal matter; 25\% dark matter;  70\% dark energy) with a similar but crucially changed version(5\% normal matter; 25\% dark matter;  70\% charged dark matter).

\bigskip

\noindent
The name dark energy was coined by Turner in 1998 shortly
after the announcement of accelerated expansion.
An outsider familiar with $E=Mc^2$ might guess that dark energy
and matter are equivalent. If our model is correct, she would
be correct although it has nothing to do with $E=mc^2$. Charged 
dark matter replaces dark energy, an ill-chosen name because it 
suggested that there exists an additional component in the Universe.

\bigskip

\noindent
In April 2024, news  from the Dark Energy 
Spectroscopic Instrument (DESI) at Kitt Peak in Arizona, USA, gave a
preliminary indication that the cosmological constant $\Lambda(t)$ 
is not constant but diminishing with time, as suggested by our Eq.(\ref{scale}),
and by our Table 1, thus providing a possible support for the EAU model.

\bigskip

\noindent
Other supporting evidence could appear in the foreseeable future
from the James Webb Space Telescope (JWST) which might shed
light on the formation of PBHs in the early universe, also from the
Vera C. Rubin Observatory in Chile which will study long duration microlensing light 
curves  which could provide evidence for the existence of PIMBHs
inside the Milky Way.

\bigskip

\noindent
It will be interesting to learn how these and other observations may
support the idea that the observed cosmic acceleration is caused
by charged dark matter.

\end{document}